\documentstyle[12pt]{article}

\newcommand{\mysection}[1]{\section{#1} \setcounter{equation}{0}}

\newcommand{\3}{\ss}

\newcommand{\be}{\begin{equation}}
\newcommand{\ee}{\end{equation}}
\newcommand{\bea}{\begin{eqnarray}}
\newcommand{\eea}{\end{eqnarray}}

\newcommand{\beas}{\begin{eqnarray*}}
\newcommand{\eeas}{\end{eqnarray*}}
\newcommand{\dis}{\displaystyle}

\newcommand{\tr}{\mbox{{\rm tr}}}
\newcommand{\zu}{\rightarrow}

\newcommand{\xv}{\vec{x}}
\newcommand{\xvp}{\vec{x}_{\perp}}
\newcommand{\yvp}{\vec{y}_{\perp}}
\newcommand{\yv}{\vec{y}}
\newcommand{\ix}{(\xv)}
\newcommand{\ixp}{(\xv_{\perp})}

\newcommand{\Ut}{\tilde{U}}
\newcommand{\Utd}{\tilde{U}^{\dagger}}
\newcommand{\Vt}{\tilde{V}}

\newcommand{\ph}{\mid \mbox{phys} \rangle}
\newcommand{\php}{\mid \mbox{phys}' \rangle}

\newcommand{\de}{\;\partial}

\newcommand{\Av}{\vec{A}}
\newcommand{\Ap}{\vec{A}_{\perp}}

\newcommand{\Piv}{\vec{\Pi}}
\newcommand{\Pip}{\vec{\Pi}_{\perp}}

\newcommand{\dev}{\vec{\de}}
\newcommand{\sv}{\vec{s}}

\newcommand{\intg}{\displaystyle \int}
\newcommand{\ointg}{\displaystyle \oint}

\newcommand{\llabel}[1]{\label{#1}
 }

\begin{document}

\begin{titlepage}
\begin{flushright}
hep-ph/9509417\\
FAU-TP3-95/9 \\
July 1995
\\
\end{flushright}
\vspace*{1.5cm}
\begin{center}% \LARGE{\bf Large and Residual Gauge Transformations \\
              %            of QCD$_{3+1}$ in an Axial Gauge}
               \LARGE{\bf Large Gauge Transformations and \\
                           Magnetic Vortices in Axial Gauge QCD}
\end{center}
\vspace*{1.0cm}
\begin{center}

{\bf Harald W. Grie\3hammer\footnote{Talk presented at the ``Workshop on
 Nonperturbative Approaches to QCD'' at the ECT*, Trento,
July 10th -- 28th, 1995. Email: hgrie@theorie3.physik.uni-erlangen.de}}

\vspace*{0.2cm}

{\em Institut f\"ur Theoretische Physik III, Universit\"at
Erlangen-N\"{u}rnberg,\\
Staudtstra\3e 7, 91058 Erlangen, Germany}
\vspace*{0.2cm}

\end{center}

\vspace*{2.0cm}

%%%%%%%%%%%%%%%%%%%%%%%%%%%%%%%%%%%%%%%%%%%%%%%%%%%%%%%%%%%%%%%%%%%%%%%%%%%%%%%
%%%%%%%%%%%%%%%%%%%%%%%%%%%%%%%%%%%%%%%%%%%%%%%%%%%%%%%%%%%%%%%%%%%%%%%%%%%%%%%

\begin{abstract}

It is shown that in the modified axial gauge version of canonically
quantized QCD$_{3+1}$ on a torus only nongeneric gauge field
configurations allow for large gauge transformations. For the other
configurations, the gauge is fixed completely.  Such configurations
carry nonzero total magnetic abelian fluxes, correspond to magnetic
vortices parallel to the coordinate axes and are incorporated using
both singular gauge fields and a change of boundary conditions.

\end{abstract}

\vskip 1.0cm
\end{titlepage}

\setcounter{page}{2}

\setcounter{footnote}{0}
\newpage

%%%%%%%%%%%%%%%%%%%%%%%%%%%%%%%%%%%%%%%%%%%%%%%%%%%%%%%%%%%%%%%%%%%%%%%%%%%%%%%
%%%%%%%%%%%%%%%%%%%%%%%%%%%%%%%%%%%%%%%%%%%%%%%%%%%%%%%%%%%%%%%%%%%%%%%%%%%%%%%

\mysection{Introduction}

 One of the main issues in nonabelian gauge theories
 is the presence of redundant variables.  Eliminating them by
 ``gauge fixing'', one hopes to identify the relevant degrees of freedom, the
 nonperturbative part of which may solve the outstanding questions in the low
 energy regime of these theories.  This hope has been fostered recently by
 lattice calculations \cite{LQCD} in the maximal Abelian gauge \cite{MAP}
 hinting on certain singular field configurations which can be interpreted as
 monopoles to be relevant for confinement and other phenomena in QCD.

 In this context, a Hamiltonian formulation of QCD is especially useful since
 it allows one to bear in mind all intuition and techniques of ordinary
 quantum mechanics; formulating the theory in terms of unconstrained,
 ``physical'' variables is the easiest way to render gauge invariant results
 in approximations. Amongst others, this has triggered interest into cases,
 eg.\ \cite{LNT}-\cite{Langetc}, in which at least a partial elimination of
 redundant degrees of freedom can be done beyond ordinary perturbation theory
 in order to obtain a deeper insight into the nonperturbative sector.

The goal of this paper is to identify and interpret physically in a quantum
mechanical framework the configurations which are connected to
nonperturbative processes, i.e. such in which the vacuum-$\vartheta$-angle is
relevant. This is done in a special completely gauge fixed formulation, namely
the modified axial gauge \cite{LNT,Yabuki} in Hamiltonian QCD on a torus $T^3$
as spatial manifold.  Here, in contradistinction to the naive axial gauge
$A_3=0$, the eigenphases of the Polyakov loop in $x_3$-direction are kept as
dynamical variables.

\vspace{2ex}

\noindent
Before proceeding, it is useful to recapitulate a general consideration.  A
gauge fixing is complete if there exists one and only one parametrization of
function space, i.e.\ if the canonical variables are uniquely determined given
a complete set of observables \cite{BW}.  A severe obstacle to a complete
elimination of redundant variables is Singer's theorem \cite{Singer} stating
that there is no local gauge fixing procedure on compact manifolds in
nonabelian gauge theories which allows for a continuous choice of exactly one
vector potential on each gauge orbit, i.e. that one will in the way of
complete gauge fixing always encounter Gribov ambiguities \cite{Gribov} for
some field configurations.

A crude, semiclassical argument goes a follows: The Pontryagin index
\cite{Nakahara}
\be
   \llabel{q}%%%%%%%%%%%%%%%%%%%%%%%%%%%%%%%%%%%%%%%%%%%%%%%%%%%%%%%%%%
    Q:= -\frac{g^2}{32 \pi^2} \int\limits_{{\cal M}\times [0;T]} d^4x\;
          \varepsilon^{\mu\nu\rho\sigma} \tr[F_{\mu\nu}(x)F_{\rho\sigma}(x)]
      =   \int\limits_{{\cal M}\times [0;T]} d^4x\;\de_{\mu} K^{\mu}(x)
\ee
is an arbitrary integer for $SU(N)$ gauge theories whenever the spatial
manifold $\cal M$ is compact and temporal development connects two identical
physical situations at times $t=0,\;T$, hence effectively compactifies
time as well. Field configurations with topological charge $Q\not=0$ are
associated with tunneling processes of finite energy and probability,
 semiclassically instantons, and are held to be important for the explanation
of many aspects of the low energy regime of QCD.

$Q$ is gauge invariant
 and can be written as the integral over a 4-divergence of a current
 $K^{\mu}(x)$ containing only fields and their derivatives \cite{Nakahara}, so
 that
\be
 Q = \oint\limits_{\de[{\cal M}\times S^1]} d^3\Sigma_{\mu} K^{\mu} -
 \oint\limits_P d^3\Sigma_{\mu}K^{\mu}\;\;.
\ee
 The second term on the right
 hand side takes into account possible singularities at points $P$, and
 $\de[{\cal M}\times S^1]$ is the boundary of the largest chart admissible on
 ${\cal M}\times S^1$.  Without loss of generality, one can choose boundary
 conditions on $\cal M$ so that the spatial integrals of the first term are
 zero. Because the gauge fixing is complete, $K^{\mu}(t=0)=K^{\mu}(t=T)$.
 Hence the current $K^{\mu}$ together with the gauge field has to be singular
 at least at one point on ${\cal M}\times S^1$ whenever $Q\not=0$.

This singularity is a coordinate singularity in the sense that its position
and nature is a priori arbitrary and depends on the gauge choosen, yet it is
indispensible for the description of underlying physics. Mathematically, there
is a close connection to the lower dimensional example of Dirac monopoles
\cite{Nakahara}, i.e.\ QED on $S^2$, where the Dirac string starting from the
centre of the sphere is seen as a singularity of the gauge field configuration
on $S^2$ which can be rotated to an arbitrary position, but cannot be
removed.  All observables (like the field energy density) remain finite, and
excluding the singularity of the Dirac string from the sphere, one arrives at
a field which is regular everywhere on the resulting disk, but has nontrivial
boundary conditions in the vicinity of the string, thus rendering nonzero
 total magnetic flux.

Therefore, in completely gauge fixed formulations it is important to
understand how these singularities occur in detail in order to
maintain them in approximations aiming at the nonperturbative sector
of QCD.

\vspace{2ex}

\noindent
In the Coulomb gauge, Jackiw, Muzinich and Rebbi \cite{JMR} have shown
in a semiclassical context the nature of the singularities and their
position at the Gribov horizon \cite{Gribov}; in a modified light cone
gauge, Franke, Novozhilov and Prokhvatilov \cite{FNP} have made
similar considerations.  Recently, Chernodub and Gubarev
\cite{Cherno} have connected instantons with the abelian monopoles in
the maximal abelian gauge \cite{MAP}.

\vspace{2ex}

\noindent
The paper is organized as follows: Section 2 will briefly review the gauge
 fixing process by unitary gauge fixing transformations (UGFT) \cite{LNOT} in
 canonically quantized QCD as described in \cite{LNT}, stressing the
 importance of the Polyakov loop and of the Jacobian arising from the
 coordinate transformation in field space. It ends with a construction of
 the residual gauge transformations in the physical Hilbert space.  Using the
 standard derivation of the vacuum-$\vartheta$-angle in the Hamiltonian
 formulation as starting point, section 3 will discuss how the boundary
 conditions have to be changed in the process of gauge fixing and for which
 field configurations large gauge transformations may occur. In section 4,
 the underlying physics is discussed, and the last section presents
 conclusions and an outlook.

%%%%%%%%%%%%%%%%%%%%%%%%%%%%%%%%%%%%%%%%%%%%%%%%%%%%%%%%%%%%%%%%%%%%%%%%%%%%%%%
%%%%%%%%%%%%%%%%%%%%%%%%%%%%%%%%%%%%%%%%%%%%%%%%%%%%%%%%%%%%%%%%%%%%%%%%%%%%%%%

\mysection{QCD Hamiltonian in the Modified Axial Gauge}

The Hamiltonian of pure QCD in the Weyl gauge $A_0=0$
\bea
  \llabel{ham}%%%%%%%%%%%%%%%%%%%%%%%%%%%%%%%%%%%%%%%%%
   H&=&\intg d^3x\;\tr[\vec{\Pi}^{2}\left(\vec{x}\right)+
   \vec{B}^{2}\left(\vec{x}\right)]   \\
  \llabel{bdef}%%%%%%%%%%%%%%%%%%%%%%%%%%%%%%%%%%%%%%%%
  B_{i}^{a}\left(\vec{x}\right)= \frac{1}{2}\epsilon_{ijk}F_{jk}^{a}
  \left(\vec{x}\right) \;\;&,&\;\;F_{kl}=\partial_{k}A_{l}-
  \partial_{l}A_{k}-ig\left[A_{k},A_{l}\right]
\eea
is quantized by imposing the canonical commutation relations
\be
   \llabel{ccr}%%%%%%%%%%%%%%%%%%%%%%%%%%%%%%%%%%%%%%%%%%%%
   \left[A_{k}^{a}\left(\vec{x}\right),\Pi_{l}^{b}\left(\vec{y}\right)\right]=
  i\delta_{kl}\delta^{ab}\delta^{(3)}\left(\vec{x}-\vec{y}\right)
\ee
between fields $\Av$ and momenta $\Piv$, where $\vec{D}= \dev -ig\Av$ is the
covariant derivative and ${\cal O}^a= 2\tr[{\cal O} t^a]$.  $t^a$ are the
$N^2-1$ hermitean, traceless generators of the Lie algebra of $SU(N)$,
$\tr[t^at^b]= \frac{1}{2}\delta^{ab}$, and $t^{a_0}$ are the $N-1$ generators
of the Cartan subalgebra, i.e. of diagonal matrices.

The Weyl gauge allows for time independent gauge transformations whose
 infinitesimal generator is Gau\3's law. It cannot be derived as an equation
of motion in the Hamiltonian formalism but has to be imposed as a constraint
on physical states $\ph$.
\be
   \llabel{gauss}%%%%%%%%%%%%%%%%%%%%%%%%%%%%%%%%%%%%%%%%%%%%%
   \left[\dev\cdot\Piv^a\ix + gf^{abc}\Av^b\ix\cdot\Piv^c\ix\right]\ph =
   0\;\;\forall a
\ee
Finite gauge transformations are implemented by the unitary operator
\bea
   \label{gbeta}%%%%%%%%%%%%%%%%%%%%%%%%%%%%%%%%%%%%%%%%%%%%%%%
    &  \Omega[\beta] := \exp-i  \intg d^3x \;2\tr\left[ - \Piv\ix\cdot
                \vec{D}\ix\beta\ix\right]\;\;,\;\; \Vt\ix=e^{ig\beta\ix}&\\
    &\Omega\left[\beta\right]\vec{A}\left(\vec{x}\right)\Omega^{\dagger}
     \left[\beta\right]
     =\Vt\left(\vec{x}\right)\left(\vec{A}\left(\vec{x}\right)+
     \dis\frac{i}{g}\dev\right)\Vt^{\dagger}\left(\vec{x}\right)
     =:{}^{\Vt}\Av\ix\;\;.& \nonumber
\eea
On a torus $T^3$, one can without loss of generality impose periodic boundary
conditions for all fields and derivatives as well as for the gauge
 transformations\footnote{Since fermions don't affect the arguments given
in this paper and only make the formulae more lengthy, I leave them out
here. Their sole trace is not to allow for twisted boundary conditions
 \cite{tHo}.}
\be
   \llabel{pbc}%%%%%%%%%%%%%%%%%%%%%%%%%%%%%%%%%%%%%%%%%%%%%%%%%%%%%%%%
    \Av(\xv^{(i)}) = \Av(\xv^{(i)}+L\vec{e}_i)\;\;,\;\;
    \Piv(\xv^{(i)}) = \Piv(\xv^{(i)}+L\vec{e}_i)\;\;,\;\;
    \Vt(\xv^{(i)}) = \Vt(\xv^{(i)}+L\vec{e}_i)
\ee
where $\xv^{(i)}$ denotes a point with vanishing $i$th component on the
 boundary of the corresponding box with length of the edge $L$.

Following \cite{LNT}, the redundant degrees of freedom can be eliminated
via an UGFT. One solves Gau\3's law for $\Pi_3\ix$ (here written symbolically)
 in the sector of physical states,
\be
   \Pi_3\ix\ph= \left[ \tilde{p}_3\ix - \frac{1}{D_3}\vec{D}_\perp\ix\cdot
   \Pip\ix\right]\ph\;\;.
\ee
Here, $\tilde{p}_3\ix$ is the zero mode part of $\Pi_3\ix$ w.r.t.\
$D_3\ix$. Performing a coordinate transformation in field space by the
unitary ``gauge fixing'' operator
\be
   U\;:\;\left\{\Av,\;\Piv\right\} \rightarrow
   \left\{\Av',\;\Piv'\;;\;\Av^{\mbox{\scriptsize unphys}},
   \;\Piv^{\mbox{\scriptsize unphys}}
  \right\}\;\;,
\ee
one induces a ``gauge transformation'' on the fields\footnote{Formally, it is
only a gauge transformation for the fields $\Ap\,,\,\Pip$ and depends on
$A_3$. The notation for $A'_3$ is slightly different form the one in
\cite{LNT}.},
\be
  \llabel{udef}%%%%%%%%%%%%%%%%%%%%%%%%%%%%%%%%%%%%%%%%%%%%%%%%%%%%%%%%%
   U\Av\ix\,U^{\dagger} = {}^{\Ut}\Av'\ix\;:\; A_3\ix= \Ut\ix\left(A_3'\ixp+
    \frac{i}{g}\de_3\right)\Utd\ix \;\;,
\ee
so that the unphysical fields $\Av^{\mbox{\scriptsize unphys}}$
don't occur in the transformed Hamiltonian
and the unphysical momenta $\Piv^{\mbox{\scriptsize unphys}}$ are eliminated
 in the transformed physical Hilbert space by the transformed Gau\3's law.

``Zero mode'' fields $A_3'\ixp$ and conjugate momenta $\Pi_3'$ obey the
 modified axial ``gauge condition'' \cite{LNT,Yabuki}
\be
   \llabel{gfix}%%%%%%%%%%%%%%%%%%%%%%%%%%%%%%%%%%%%%%%%%%%%%%%%%%%%%
    A_3'\ixp\;,\;\Pi_3'\ixp\mbox{ diagonal}\;\;,\;\;\de_3
    A_3'\ixp= 0=\de_3\Pi'_3\ixp
\ee
and remain relevant degrees of freedom since $A'_3\ixp$ are the phases of the
 gauge invariant eigenvalues $\exp igLA_3'\ixp$ of the Polyakov loop
\be
   \llabel{polyakov}%%%%%%%%%%%%%%%%%%%%%%%%%%%%%%%%%%%%%%%%%%%%%%
   \mbox{P}\!\exp ig \int\limits_0^L dx_3\;A_3\ix \;\;.
\ee
This means that transverse, colour neutral gluons moving in the
 $(x_1,x_2)$-plane with polarization in the $x_3$-direction remain physical.
 It is also expressed in the fact that the solution to (\ref{udef}) cannot be
 given for $A_3'\ix=0$ since then $\Ut$ would not be periodic in
 $x_3$-direction.  Allowing for a colour neutral zero mode, one finds
\be
   \llabel{solnut}%%%%%%%%%%%%%%%%%%%%%%%%%%%%%%%%%%%%%%%%%%%%%%%%
   \Ut\ix= \mbox{P}e^{ig \int\limits_0^{x_3} dy_3\;A_3(\xvp,y_3)}
	 e^{ig\Delta\ixp} e^{-igx_3 A_3'\ixp}\;\;,
\ee
where $e^{ig\Delta\ixp}$ diagonalizes the Polyakov loop (\ref{polyakov})
\be
   \llabel{diag}%%%%%%%%%%%%%%%%%%%%%%%%%%%%%%%%%%%%%%%%%%%%%%%%%%%
   e^{ig\Delta\ixp} e^{igL A_3'\ixp}e^{-ig\Delta\ixp}=\mbox{P}\!\exp ig
   \int\limits_0^L dx_3\;A_3\ix\;\;.
\ee
In the space of transformed physical states
\be
   \llabel{trafoph}%%%%%%%%%%%%%%%%%%%%%%%%%%%%%%%%%%%%%%%
   \mid \mbox{phys}'[A'] \rangle:=\sqrt{{\cal J}[A'_3]}\,U[A_3]
   \mid \mbox{phys} [A]\rangle\;\;,
\ee
the resulting Hamiltonian and canonical commutation relations are \cite{LNT}
\be
   \llabel{hamiltonian}%%%%%%%%%%%%%%%%%%%%%%%%%%%%%%%%%%%%%%%%%%%%
   \begin{array}{c}
    \langle\mbox{phys}'_1|H'|\mbox{phys}'_2\rangle= \langle\mbox{phys}'_1|
   \Bigg\{\intg d^3x\; \tr[\Piv'^2\ix +  \vec{B}'^2\ix] + \vspace*{\jot}\\
   \!\!\!+ \intg d^2x_{\perp}\!\!\intg
   dx_3\!\intg dy_3\!\!\!\!\dis\sum\limits_{{pq\,n}\atop {p\not=q\forall n=0}}
   {\dis \frac{G'_{\perp qp}(\xvp,x_3) G'_{\perp pq}(\xvp,y_3)}{\left[
   \frac{2\pi}{L}n+g\left(A'_{3,q}\ixp - A'_{3,p}\ixp\right)\right]^2}}
   e^{\frac{2\pi i}{L} n(x_3-y_3)}\Bigg\}|\mbox{phys}'_2\rangle
  \end{array}
\ee
\be
   \llabel{ccrels}%%%%%%%%%%%%%%%%%%%%%%%%%%%%%%%%%%%%%%%%%%%%%%%%
   \left[A'^a_i\ix,\Pi'^b_j\ix\right] = \left\{
         \begin{array}{lcl} i\delta_{ij}\delta^{ab}\delta^{(3)}(\xv-\yv)
            &\mbox{for}& i=1,2\\
            i\delta_{ij}\delta^{ab}\delta^{(2)}(\xvp-\yvp)&\mbox{for}&i=3
                          \end{array} \right.\;\;.
\ee
$\vec{B}'\ix$ is (\ref{bdef}) with primed replacing unprimed variables
as in $G'_{\perp}\ix := \vec{D}_{\perp}'\ix\cdot\Pip'\ix$, and the Green's
 function of $D_3$ has been given explicitely in the ``Coulomb'' part,
 i.e.\ the last line of $H$. $G'_{\perp pq}$ is
 the $(pq)$ entry of the matrix $G'_{\perp}$, and $A'_{3,p}$  the
$(pp)$ entry of the diagonal matrix $A'_3$. The sum goes over all labels
$(pq\,n)$ for which the denominator is nonzero. In (\ref{trafoph})
it was taken into account that a change of variables induces a Jacobian ${\cal
 J}[A'_3]$
in the Hilbert space measure\footnote{A normalization constant due to the
integration over unphysical degrees of freedom has been dropped.}
\bea
   \int {\cal D}A_3 \rightarrow \int{\cal D} A'_3 \;{\cal J}[A'_3]&=&
   \int {\cal D}A'_3 \;\exp\delta^{(2)}(\vec{0}_{\perp})\int d^2x_{\perp}
    \ln J[A'_3\ixp]\;\;,\nonumber \\
   \llabel{jac}%%%%%%%%%%%%%%%%%%%%%%%%%%%%%%%%%%%
    {\cal J}[A'_3]=\prod\limits_{\xvp}J[A'_3\ixp]&\!\!,&\!\!
   J[A'_3\ixp] = \prod\limits_{p>q} \sin^2\frac{gL}{2}\left[ A'_{3,q}\ixp-
                    A'_{3,p}\ixp\right] \;,
\eea
where $ J[A'_3\ixp]$ is the Haar measure of $SU(N)$ for the case that the
integrand depends only on the invariants. The measure in the Schr\"odinger
 representation of the Hilbert space
of ``radial wave functions'' $\php$ is again $\int {\cal D}A'$, and $\php$
vanishes at the zeroes of the Jacobian,
\be
  \llabel{radwave}%%%%%%%%%%%%%%%%%%%%%%%%%%%%%%%%
  \php = 0 \;\forall \;{\cal J}[A'_3]=0 \; \mbox{, i.e. }
  \frac{gL}{2\pi}\left[ A'_{3,q}\ix-A'_{3,p}\ixp \right]\in Z \mbox{ for some }
  \xvp\;,p\not= q\;.
\ee
The importance of this boundary condition in field space has been stressed
recently \cite{Lenzoffs} and is a remnant of the fact that $A'_3\ixp$ is an
angle variable whose extraction from the Polyakov loop (\ref{polyakov}) is not
unique.

Because of the occurrence of the zero modes $A'_3,\Pi'_3$, a residual Gau\3's
law
\be
   \llabel{resglaw}%%%%%%%%%%%%%%%%%%%%%%%%%%%%
    \int dx_3 \; G'^{a_0}_{\perp}\ix \php = 0 \;\;\forall a_0
\ee
survives which can be interpreted as eliminating all colour neutral,
longitudinal gluons moving in the $(x_1,x_2)$-plane. It can be solved as above
 \cite{LNT}, but this is not necessary in what follows.

\vspace{2ex}

\noindent
The residual gauge transformations can either be constructed by explicit
transformation of (\ref{gbeta}) to the transformed physical Hilbert space
 \cite{hgrie}
\be
   \Omega'[\beta]\php = \sqrt{{\cal J}[A'_3]} \;U \Omega[\beta] U^{\dagger}
                        \frac{1}{\sqrt{{\cal J}[A'_3]}}\php\;\;,
\ee
or -- as sketched here --  by an inspection \cite{LNT} of the freedoms in the
 solution (\ref{solnut}) of equation (\ref{udef}) defining $A'_3\ixp$, the
 eigenphases of the Polyakov loop (\ref{polyakov}).

$\Omega'[\beta]\Av'\ix\Omega'^{\dagger}[\beta]$ again has to obey the gauge
condition (\ref{gfix}), so that the freedom to perform gauge transformations
which mix diagonal and offdiagonal components of the fields $\Av'$ is removed.
Together with the demand for periodicity of the gauge transformations
(\ref{pbc}), this yields in the generic case
\be
  \llabel{resgtrafo}%%%%%%%%%%%%%%%%%%%%%%%%%%%%%%%
  \Omega'[\beta]\Av'\ix\Omega'^{\dagger}[\beta]= {}^{\Vt'}\Av'\ix\;\;:\;\;
  \Vt' = R \exp i\left[\frac{2\pi }{L}\vec{n}\cdot \xv+\beta_{\mbox{\scriptsize
  period}}\ixp\right]\;\;.
\ee
$\beta_{\mbox{\scriptsize period}}\ixp$ are arbitrary gauge functions periodic
 in $\xvp$ in a traceless diagonal matrix, $\vec{n}$ a vector consisting of
 diagonal traceless matrices having integer entries, and $R$ a member of the
 $N$ dimensional representation of the permutation group ${\cal S}_N$, the
 Weyl (reflection) group of $SU(N)$, which changes the order of the entries
 in a diagonal matrix. The residual gauge group is therefore
\be
   \llabel{resggroup}%%%%%%%%%%%%%%%%%%%%%%%%%%%%%%%%%%%%%%%%%%%%%%%%%%%%
   G':=\left[U(1)\right]^{N-1}\times {\cal S}_N\;\;,
\ee
 and (as can be shown explicitely \cite{hgrie})
\be
   \llabel{resomega}%%%%%%%%%%%%%%%%%%%%%%%%%%%%%%%%%%%%%%%
   \Omega'[\beta] =  \exp \Bigg\{-i\int d^3x\; 2\tr[-\frac{2\pi}{gL} \left(
    \Piv'\cdot
    \vec{D}' \right)\left(\vec{n}\cdot \xv\right) + \left( \vec{D}'\cdot\Piv'
    \right)\beta_{\mbox{\scriptsize period}}\ixp]\Bigg\}\;\Omega_R\;\;,
\ee
where $\Omega_R$ is the operator generating the Weyl permutations.

When the Polyakov loop has degenerate eigenvalues on some line
$(\xv_{\perp\,0},[0;L])$, the residual gauge group is larger on this closed
loop in the $x_3$-direction, namely $G'=
\left[U(1)\right]^{m-1}\otimes_{i=1}^{m-1} SU(\alpha_i)\times {\cal S}_N$
\cite{FNP,GPY} for $m$ different eigenvalues, each of degeneracy
$\alpha_i$. At these points, the Jacobian (\ref{jac}) of the coordinate
transformation in field space vanishes, and hence the gauge fixing procedure
is not defined at all. Sometimes, it is argued \cite{Langetc} that these
nongeneric points in configuration space are mere coordinate singularities and
don't have any gauge invariant meaning, especially since their set has zero
measure and is in addition suppressed because the Coulomb part in the
Hamiltonian (\ref{hamiltonian}) yields infinite energy. The arguments given
in the introduction show that this may indeed be misleading and not render the
properties of low-energy QCD known: Neglecting such configurations in a
completely gauge field formulation, no winding number changing processes can be
described.

One may nonetheless perform the gauge fixing in the above way for all points
 on $T^3$ for which the transformation $\Ut\ix$ is single-valued and the
 Jacobian (\ref{jac}) is nonzero. All integrals are then understood to
exclude these singular loops and so go over a manifold $T^3_R$ which may have
holes. Then (\ref{resgtrafo}/\ref{resomega}) gives the residual gauge
transformations on $T^3_R=T^2_R\times S^1_{x_3}$, of which the displacements
 of the colour neutral fields
\be
   \llabel{displacements}%%%%%%%%%%%%%%%%%%%%%%%%%%%%%%%%%%
   \Omega'[\beta]\Av'_p\ix\Omega'^{\dagger}[\beta] =
   \Av'_p\ix+\frac{2\pi}{gL}\vec{n}\;\;,\;\;
   \Omega'[\beta]\Piv'_p\ix\Omega'^{\dagger}[\beta] = \Piv'_p\ix \nonumber
\ee
affect only their zero modes and will be of importance in what follows.
 Offdiagonal fields are rotated by $\Vt'$.

The freedom to perform periodic, colour neutral gauge transformations
in $\xvp$-direction is void in the transformed physical Hilbert space because
of the residual Gau\3's law (\ref{resglaw}), cf.\ (\ref{resomega}).
 Note finally that the modified
axial gauge lies in the class of abelian projection gauges \cite{MAP}: The
nonabelian $SU(N)$ gauge freedom is reduced to leave only $N-1$
 ``electrodynamics''.

%%%%%%%%%%%%%%%%%%%%%%%%%%%%%%%%%%%%%%%%%%%%%%%%%%%%%%%%%%%%%%%%%%%%%%%%%%%%%%%
%%%%%%%%%%%%%%%%%%%%%%%%%%%%%%%%%%%%%%%%%%%%%%%%%%%%%%%%%%%%%%%%%%%%%%%%%%%%%%%

\mysection{Large and Residual Gauge Transformations}

Before gauge fixing, the following quantum mechanical
derivation for the occurrence of the vacuum-$\vartheta$-angle can be given
\cite{JRCDG}.

Physical states are annihilated by Gau\3's law, the generator of
infinitesimal gauge transformations,  and hence are invariant under  small
 gauge transformations (winding number zero). The integral over the
Chern--Simons three-form (zero component of the vector $K^{\mu}$ (\ref{q})),
\be
  \llabel{wa}%%%%%%%%%%%%%%%%%%%%%%%%%%%%%%%%%%%%%%%%%%%%%%%%%%
       W[A] := \frac{g^2}{16 \pi^2} \intg d^3x \;\epsilon^{ijk}
	     \tr[ F_{ij}A_k + \frac{2i}{3}g A_iA_jA_k]\;\;,
\ee
serves as detector for the winding number of a gauge transformation,
distinguishing small and large ones. It commutes with Gau\3's law and
therefore is invariant under small, but changes under large ones
\bea
    \llabel{trafowa}%%%%%%%%%%%%%%%%%%%%%%%%%%%%%%%%%%%%%%%%
    \Omega[\beta] W[A] \Omega^{\dag}[\beta] &=& W[A] + \nu(\Vt)
              +\frac{ig}{8\pi^2}
	    \intg d^3x \;\epsilon^{ijk} \de_i \tr[A_j(\de_k\Vt^{\dag})\Vt]
\eea
by the (additive) winding number
\bea
   \llabel{defn}%%%%%%%%%%%%%%%%%%%%%%%%%%%%%%%%%%%%%%%%%%%%
    \nu(\Vt) &:=& \frac{1}{24\pi^2} \intg d^3x \;\epsilon^{ijk}
	    \tr[(\Vt\de_i\Vt^{\dag})(\Vt\de_j\Vt^{\dag})(\Vt\de_k\Vt^{\dag})]
            \;\in Z\;\;.
\eea
The surface term in (\ref{trafowa}) vanishes because of the periodicity
condition  (\ref{pbc}). This is closely related to the vanishing of all the
total colour charges in the box \cite{hgrie,Bala}
\be
    \llabel{charges}%%%%%%%%%%%%%%%%%%%%%%%%%%%%%%%%%%%%%%%%%%%%%%%%%%%%%
    Q^a\ph:=\intg d^3x\; \Big( f^{abc} \Av^b\ix\cdot\Piv^c\ix \Big)\ph = 0
    \;\; \forall a\;\;.
\ee
Since $\Omega[\beta] W[A] \Omega^{\dag}[\beta]$ is in the physical Hilbert
sector,
\bea
   \llabel{elimsurf}%%%%%%%%%%%%%%%%%%%%%%%%%%%%%%%%%%%%%%%%%%
   & \Big[Q^a,\Big[Q^a,\Omega[\beta] W[A] \Omega^{\dag}[\beta]\Big]\Big]\ph
  \stackrel{!}{=}   0 = {\dis\frac{ig}{8\pi^2}}
  \intg d^3x \;\epsilon^{ijk} \de_i\tr[A_j(\de_k\Vt^{\dag})\Vt]\ph\;\;.&
\eea
$\Omega[\beta]$ therefore leaves physical states invariant only up to a phase
into which -- besides the winding number -- the famous vacuum-$\vartheta$-angle
 enters as new, hidden parameter:
\be
   \Omega[\beta]\ph= e^{i\vartheta\nu(\Vt)}\ph
\ee

\vspace{2ex}

\noindent
If one assumed that all residual gauge transformations after the elimination
of Gau\3's law were large, one would require from (\ref{resomega})
\be
   \Omega'[\beta]\php = e^{i\vec{\vartheta}\cdot\vec{n}}\php
\ee
where $\vec{\vartheta}$ are $3(N-1)$ independent parameters in a vector of
diagonal traceless matrices. This is obviously in conflict with the occurrence
of only {\em one} free parameter before gauge fixing. Large and residual
gauge transformations can therefore not be identical.

The correct procedure to identify large gauge transformations after gauge
fixing is to start from the winding number detector
(\ref{wa}) in the modified axial gauge, where $\Av$ is replaced by $\Av'$,
and investigate under which of the residual gauge transformations it changes
by an integer:
\be
   \llabel{trafowap}%%%%%%%%%%%%%%%%%%%%%%%%%%%%%%%%%%%%%%%%%
  \begin{array}{rcl}
  \Omega'[\beta] W[A'] \Omega'^{\dag}[\beta] &\stackrel{!}{=}& W[A']+\nu(\Vt)\\
   &=&W[A'] + n(\Vt') +\dis\frac{ig}{8\pi^2}\int\limits_{T^3_R} d^3x
            \;\epsilon^{ijk} \de_i \tr[A'_j(\de_k\Vt'^{\dag})\Vt']
  \end{array}
\ee
Since $\dev\Vt'$ is abelian ($\dev R=0$) in $T^3_R$, $n(\Vt')=0$ and hence the
 surface integral must be nonvanishing, which suggests that regularity or
 boundary conditions of the fields have to be changed by the UGFT.

A valuable clue towards an identification of the relevant degrees of freedom
yielding nonzero winding number is the fact that in the transformed physical
Hilbert space, only the neutral components of the transformed total colour
charges annihilate physical states
\be
   \llabel{qphys}%%%%%%%%%%%%%%%%%%%%%%%%%%%%%%%%
    Q'^{a_0} \php :=
   \int\limits_{T^3_R} d^3x\; \left[  f^{a_0bc} \Ap'^b\ix \cdot
            \Pip'^c\ix \right] \php \stackrel{!}{=} 0 \;\;\forall a_0\;\;,
\ee
since offdiagonal global transformations in general violate the gauge
condition (\ref{gfix}). Indeed, condition (\ref{qphys}) can also be derived
by performing the UGFT of (\ref{charges}) \cite{hgrie} or by integrating
(\ref{resglaw}) over $\xvp$ and using periodicity and continuity for the
longitudinal colour neutral fields $\Av'^l_{\perp,p}$ which will later be
 demonstrated to hold.

The analogous calculation to (\ref{elimsurf}) eliminates only the offdiagonal
components of the surface integral
\be
   \llabel{surfq}%%%%%%%%%%%%%%%%%%%%%%%%%%%%%%%%%%%%%%%%%%%
       \int\limits_{T^3_R} d^3x\; \epsilon^{ijk} \de_i
   \tr[A_j'(\de_k\tilde{V}'^{\dag})\tilde{V}']\php = \frac{1}{2}
    \int\limits_{T^3_R} d^3x \;\epsilon^{ijk} \de_i
    \Big[A_j'^{a_0}[(\de_k\tilde{V}'^{\dag})\tilde{V}']^{a_0}\Big]\php
\ee
and hence suggests to investigate boundary conditions and continuity
properties of the colour neutral fields after the UGFT in order to identify the
 large gauge transformations.

\vspace{2ex}

\noindent
All variables are still periodic in $x_3$ since $\Ut$(\ref{solnut}) is
 explicitely periodic and continuous if $A_3\ix$ was. On the other hand,
periodicity in $\xvp$ can in general not be maintained for two reasons.

Although the eigenvalues $\exp igLA_3'\ixp$ of the Polyakov loop
(\ref{polyakov}) are periodic and continuous, the last term in the explicit
solution $\Ut$ (\ref{solnut}) makes it necessary to take its logarithm in
order to define $A'_3\ixp$. As has been demonstrated in \cite{LNT},
$A'_3\ixp$ may be chosen to be continuous on all of $T^3$, but then the
boundary conditions of the phases will in general be changed to\footnote{
This definition of $m_i$ differs from the one given in \cite{LNT}.}
\be
   A'_3(\vec{x}^{\,(i)}+L\vec{e}_{i})-A'_3(\vec{x}^{\,(i)})= \frac{2\pi}{gL}
   \varepsilon^{ij}m_j\;\;\mbox{ for }  i=1, 2
\ee
where $\varepsilon^{ij}$ is the totally antisymmetric unit tensor of second
 rank, $i, j=1, 2$ and $m_i$  a diagonal, traceless matrix with integer
 entries. The mapping $\exp igLA_3'\ixp\;:\;T^2_R\zu \left[U(1)\right]^{N-1}$
 decomposes into topologically
distinct classes labelled by the two winding numbers $m_i\in Z^{N-1}$.

Furthermore, the diagonalization matrix $e^{ig\Delta}$ (\ref{diag}) is
determined up to right multiplication with an element of the equivalence
class of $\exp igLA_3'\ixp$ \cite{FNP,GPY}, i.e. $ e^{ig\Delta\ixp} \in
SU(N)/G'$ ($\xvp \in T^2_R$) lies in the coset. This suggests the
occurrence of additional singularities in the transverse fields $\Ap$
whenever $(\de_1\de_2-\de_2\de_1)e^{ig\Delta\ixp}\not= 0$. Nonetheless,
such points
have been excluded by the choice of $T^3_R$. The diagonalization can
be chosen to be continuous on $T^3_R$, but in general will not be periodic
because the mapping $SU(N)/G'\to T^2_R$ decomposes into topologically distinct
classes or the first homotopy group of $\left[U(1)\right]^{N-1}$ is $Z^{N-1}$.
\be
   e^{ig\Delta(\vec{x}^{\,(i)}+L\vec{e}_{i})}=e^{ig\Delta(\vec{x}^{\,(i)})}
    h(\vec{x}^{\,(i)})\;\;,\;\; h(\vec{x}^{\,(i)})\in G'
\ee
So it proves again impossible to find a gauge fixing procedure to the
modified axial gauge (\ref{gfix}) which is both periodic and continuous for all
gauge field configurations. If one wants to preserve continuity, the boundary
conditions change from (\ref{pbc}) to
\bea
    &U\left( \Av(\vec{x}^{\,(i)}+L\vec{e}_{i}) - \Av(\vec{x}^{\,(i)})\right)
    U^{\dagger}=  {}^{\Ut}\Av'(\vec{x}^{\,(i)}+L\vec{e}_{i}) -
    {}^{\Ut}\Av'(\vec{x}^{\,(i)})= 0& \nonumber\\
  \llabel{newbc}%%%%%%%%%%%%%%%%%%%%%%%%%%%%%%%%%%%%%%%%%%%%%%%%%%%%%%
  &\Longrightarrow \Av'(\vec{x}^{\,(i)}+L\vec{e}_{i})=
    u^{(i)}\left(\Av'(\vec{x}^{\,(i)}) +\frac{i}{g}\dev\right)u^{(i)\dagger}&\\
  &\Piv'(\vec{x}^{\,(i)}+L\vec{e}_{i})=u^{(i)}
  \Piv'(\vec{x}^{\,(i)})u^{(i)\dagger}\;\;,\;\;
   \Vt(\vec{x}^{\,(i)}+L\vec{e}_{i})=u^{(i)}
   \Vt(\vec{x}^{\,(i)})u^{(i)\dagger}\;\;,&\nonumber
\eea
where
\bea
   u^{(i)}(\vec{x}^{\,(i)})=\left\{ \begin{array}{ccl}
            e^{\frac{2\pi i}{L}x_3\varepsilon^{ij}m_j} h(\vec{x}^{\,(i)})
                    & \mbox{for}& i=1, 2 \\
            {\bf 1} & \mbox{for}&  i=3 \end{array}\right. \;\;.
\eea
This shows that both the residual gauge transformations and the longitudinal
colour neutral fields in (\ref{resglaw}) again have to be periodic
and continuous, justifying (\ref{resgtrafo}) and (\ref{qphys}), while the
transverse colour neutral fields are periodic up to a shift, as the rest of the
fields up to a rotation about axes in colour space corresponding to diagonal
 generators.

The winding numbers of the above mappings  do not interfere since different
diagonalization matrices $e^{ig\Delta}$ cannot change the eigenvalues of the
Polyakov loop and vice versa and are the diagonal matrices
\bea
     & m_1 = \dis\frac{g}{2\pi}\intg\limits_{T^2_R} dx_3\,dx_2\;\de_2
     A'_{3,p}\ixp=
     \frac{g}{2\pi L}\intg\limits_{T^3_R} d^3x\;B_{1,p}'\ix &\nonumber\\
 \llabel{windingno}%%%%%%%%%%%%%%%%%%%%%%%%%%%%%%%%%%%%%%
    &m_2 = -\dis\frac{g}{2\pi}\intg\limits_{T^2_R} dx_3\,dx_1\;\de_1
    A'_{3,p}\ixp=
     \frac{g}{2\pi L}\intg\limits_{T^3_R} d^3x\;B_{2,p}'\ix&\\
    &m_3 = \dis\frac{g}{2\pi}\ointg\limits_{\de[T^2_R]} d\sv_{\perp}\cdot
          \Av'_{\perp,p}\ix - \frac{g}{2\pi}\ointg\limits_{P}
           d\sv_{\perp}\cdot\Av'_{\perp,p}\ix
       =\frac{g}{2\pi L}\intg\limits_{T^3_R} d^3x\;b_{3,p}'\ix\;\;.& \nonumber
\eea
(Note that $b'_{3,p}:=\de_1A'_{2,p}-\de_2A'_{1,p}\not= B'_{3,p}$ and
$\tr[\vec{m}]=0$.) $\vec{m}$ is invariant under residual gauge transformations
(\ref{resgtrafo}), except for permutation of entries, and hence an observable.
Together with (\ref{trafowap}/\ref{surfq}), this shows that only
configurations  with nonzero (abelian) magnetic fluxes through the box can
 have mirror configurations which differ by a large gauge transformation of
 winding number
\be
  \llabel{nup}%%%%%%%%%%%%%%%%%%%%%%%%%%%%%%%%%%%
   \nu(\Vt')= \frac{1}{2}\tr[\vec{n}\cdot\vec{m}] \;\;.
\ee
These configurations describe abelian magnetic vortices whose total magnetic
 flux obeys the Dirac quantization condition. This is again reminiscent of the
 Dirac monopole as discussed in the introduction.

 With respect to all other gauge configurations, the gauge is therefore
completely fixed (up to small gauge transformations connected to the residual
Gau\3's law (\ref{resglaw})). The opportunity not only for small,
but also for large gauge transformations has been eliminated for all zero
magnetic flux configurations in the modified axial gauge (\ref{gfix}).

%%%%%%%%%%%%%%%%%%%%%%%%%%%%%%%%%%%%%%%%%%%%%%%%%%%%%%%%%%%%%%%%%%%%%%%%%%%%%%%
%%%%%%%%%%%%%%%%%%%%%%%%%%%%%%%%%%%%%%%%%%%%%%%%%%%%%%%%%%%%%%%%%%%%%%%%%%%%%%%

\mysection{Discussion}

That the resolution of Gau\3's law in the space of transformed physical states
eliminated indeed not only the opportunity for small, but also for large gauge
transformations for many configurations, can be traced back to the fact that
in general the UGFT will not leave $W[A]$ (\ref{wa}) invariant, but changes it
by
\be
   U\,W[A]\,U^{\dagger} = W[A'] + \nu(\Ut) +\frac{ig}{8\pi^2}
	    \intg d^3x \;\epsilon^{ijk} \de_i \tr[A_j(\de_k\Ut^{\dag})\Ut]\;\;.
\ee
One should note that $\Ut[A_3]$ depends on the unphysical variables
whose conjugate momenta have been eliminated in the physical Hilbert space.
One can show \cite{hgrie} that there exists indeed no solution $\Ut[A_3]$ to
the gauge fixing procedure (\ref{udef}/\ref{gfix}) which is periodic and
continuous in all directions, as well as ``small'', i.e.\ which would yield
\be
  \llabel{triv}%%%%%%%%%%%%%%%%%%%%%%%%%%%%%%%%%%%%%%%%%%%%%%%%%%%%
  U\,W[A]\,U^{\dagger} = W[A']
\ee
for all field configurations. If (\ref{triv}) would hold, every point in the
physical Hilbert space had mirror points of the same physics which can be
reached by large gauge transformations. This is the case for the axial and
 Coulomb gauge representation of QED \cite{LNOT}, where the UGFT leaves the
zero modes of the fields unchanged, which are
the winding number detectors. Therefore with the resolution of Gau\3's law,
only the unitary operator of small gauge transformations is reduced to the
identity in the physical sector. In contradistinction, in modified axial gauge
QCD one has to allow for singular or nonperiodic gauge configurations to
 implement large gauge transformations.

Although the set they form may be of zero measure, these configurations are
indispensible since only their occurrence can be connected to the existence of
the vacuum-$\vartheta$-angle which is known to be relevant at least in
semiclassical approximations.

\vspace{2ex}

\noindent
On the other hand, several questions need clarification.  Since with
(\ref{hamiltonian}/\ref{windingno}/\ref{newbc}), $[H',\vec{m}_{\perp}]=0$, the
total magnetic fluxes in $\xvp$-directions are constants of motion.  Choosing
furthermore the fields to be continuous in $T^3$, the vortex configurations
are described by fields $A'_3$ which contain at least $2\vec{m}_{\perp}$
closed loops in $x_3$-direction at which the Jacobian ${\cal J}[A'_3]$
(\ref{jac}) is zero, yielding zero wavefunction (\ref{radwave}).  One may
therefore rule out tunneling processes between configurations with
$\vec{m}_{\perp}\not=0$, which are connected by displacements
$\vec{n}_{\perp}\not=0$, although they are connected by large gauge
transformations (\ref{nup}), and restrict oneself to the only sector of
nonzero probability amplitude, $\vec{m}_{\perp}=0$.

The argument is less stringent for $m_3$ since -- $[H',m_3]$ being nonzero --
 it is not time independent. Still, its equation of motion is undecided
 whenever the denominator of the ``Coulomb term'' in (\ref{hamiltonian})
 becomes zero in time evolution.  If time evolution between two configurations
 connected by displacements $n_3\not=0$ is again continuous for $A'_3$, then
 it crosses -- as above -- at $2n_3$ points in time configurations with zero
 Jacobian.  At these points, the (radial) wavefunction vanishes, which
 seemingly again forbids tunneling between
 $m_3\not=0\,,\,n_3\not=0$-configurations despite of their being partly
 connected by large gauge transformations.

One might therefore be tempted to conclude that -- although large gauge
transformations exist -- the vacuum-$\vartheta$-angle becomes irrelevant
because every (semiclassical) tunneling process between configurations
connected by large gauge transformations is suppressed by the Jacobian. On the
other hand, Singer \cite{Singer} shows that the points in configuration space
at which a complete gauge fixing procedure is singular (i.e.\ at the Gribov
horizon) will necessarily have zero Jacobian. The Gribov horizon is then just
 defined as this manifold of gauge configurations which forms an impenetrable
 barrier for field configurations in time evolution. Only there, an
identification of points which differ by a large gauge transformation is
 possible and incorporates the physics connected to the
 vacuum-$\vartheta$-angle. This is reminiscent of the picture of a pendulum at
turning point \cite{BW,JMR}.

It is well known that before gauge fixing there exists a semi-classical
configuration (the instanton) of finite action and energy which extrapolates
between physically equivalent configurations connected by large gauge
transformations for every winding number. Such semiclassical solutions to the
Heisenberg equations of motion resulting from (\ref{hamiltonian}) must be
found after the UGFT, too, now running between two points at the Gribov
Horizon.  Maybe dynamics indeed forbids one to interpolate in time between
certain configurations $\vec{m}\not=0\,,\,\vec{n}\not=0$. Such a decoupling
would also provide a solution to the problem of half-integer winding numbers
occuring in (\ref{nup}) for $SU(N)\,,\,n>2$, whenever $\vec{m}$ and $\vec{n}$
are members of overlapping, but not identical subalgebrae of the Cartan
algebra.

\mysection{Conclusions}

Gauge fixing to the modified axial gauge, in which the eigenphases of the
Polyakov loops in $x_3$-direction are kept as dynamical variables, yielded
with the Hamiltonian (\ref{hamiltonian}) unique vector potentials at the
generic points in the physical sector of configuration space. The {\em
non}generic points are defined as the ones at which the coordinate
transformation (\ref{udef}) in field space which is part of the gauge fixing
process (UGFT) \cite{LNT,LNOT} becomes singular or non-periodic in $\xvp$.
The unitary operator $\Omega'$ (\ref{resomega}) implementing residual gauge
transformations $\Vt'$ (\ref{resgtrafo}) has been given in the physical
Hilbert space, i.e.\ in terms of unconstrained variables.  Large gauge
transformations exist only for the nongeneric set of configurations.

The winding number of a residual gauge transformation was -- due to these
singularities -- not given by a field-independent volume integral (\ref{defn})
but by a surface integral (\ref{trafowap}/\ref{surfq}/\ref{nup}) which
measured the total colour neutral magnetic fluxes (\ref{windingno}) in the
directions in which the displacements of the zero modes of the colour neutral
gluons (\ref{displacements}) act on the fields.  Therefore, the existence of
abelian magnetic vortex configurations was necessary to allow for large gauge
transformations in the modified axial gauge.

The occurrence of singularities and changes of boundary conditions proved to
 be a natural outcome of the UGFT (\ref{udef}/\ref{solnut}).  Besides
 ambiguities in the diagonalization of the Polyakov loop which had already
 been found in \cite{FNP,GPY}, this was attributed to a projection of its
 eigenphases  $A'_3\ixp$ at the boundaries of the box onto different Riemann
 sheets \cite{LNT}.

Vortex configurations with nonzero magnetic flux in the $\xvp$-directions
are connected to the problems in extracting the eigenphases. Taking into
account the dynamics of the quantum mechanical system, it seems that their
wave function is zero because of the Jacobian (\ref{jac}/\ref{radwave})
arising from the coordinate transformation of the UGFT to curvilinear
coordinates in the physical Hilbert space. Nonetheless, the
vortex configurations with nonzero magetic flux in $x_3$-direction, arising
from ambiguities in the diagonalization of the Polyakov loop, are
admissible. For them, time evolution between configurations differing by a
large gauge transformation seems to be forbidden, again because of the
Jacobian. That it appears to prohibit any tunneling, is a quantum mechanical
phenomenon and lies beyond semiclassical treatment. On the other hand, points
at which the Jacobian is zero were excluded from $T^3$ in the UGFT, and a
semiclassical approximation should render the well-known results of instanton
calculations.

\vspace{2ex}

\noindent
A detailed, more physically motivated study of the winding number changing
processes and their possible suppression by (or circumvention of) the Jacobian
is under way. The question to what extend the magnetic vortices are localized
in a semiclassical solution of the dynamics is presently addressed, too. It is
also investigated whether these vortices may serve as relevant configurations
for the description of low energy properties of QCD, as in the abelian
projection gauges \cite{LQCD,MAP}. The hope is that -- having understood these
issues -- one
may perform suitable approximations in the completely gauge fixed Hamiltonian
framework in order to improve our insight into the low energy sector of QCD,
which is generally attributed to topological ``nontrivialities''.

%%%%%%%%%%%%%%%%%%%%%%%%%%%%%%%%%%%%%%%%%%%%%%%%%%%%%%%%%%%%%%%%%%%%%%%%%%%%%%%
%%%%%%%%%%%%%%%%%%%%%%%%%%%%%%%%%%%%%%%%%%%%%%%%%%%%%%%%%%%%%%%%%%%%%%%%%%%%%%%
\vspace{2ex}

\begin{center}
{\bf Acknowledgment}
\end{center}
I thank P.\ van Baal, F.\ Lenz, H.\ W.\ L.\ Naus, J.\ Polonyi and J.\ J.\ M.\
Verbaarschot for valuable discussions, as well as D.\ Lehmann, J.\ Levelt
 and D.\ Stoll for everlasting support and T.\ Kraus and M.\ Seeger
for a careful reading of the manuscript. I am very grateful to D.\ Diakonov
 for organizing this fantastic workshop, and to all its participants.

%%%%%%%%%%%%%%%%%%%%%%%%%%%%%%%%%%%%%%%%%%%%%%%%%%%%%%%%%%%%%%%%%%%%%%%%%%%%%%%
%%%%%%%%%%%%%%%%%%%%%%%%%%%%%%%%%%%%%%%%%%%%%%%%%%%%%%%%%%%%%%%%%%%%%%%%%%%%%%
%%%%%%%%%%%%%%%%%%%%%%%%%%%%%%%%%%%%%%%%%%%%%%%%%%%%%%%%%%%%%%%%%%%%%%%%%%%%%%


\begin{thebibliography}{99}

\bibitem{LQCD} L.\ Del Debbio, A.\  Di Giacomo and G.\ Paffuti, {\em
		 Phys.\ Lett.\ }{\bf B355}(1995), 255;\\
   		 L.\ Del Debbio, A.\  Di Giacomo and G.\ Paffuti, {\em
		 Nucl.\ Phys.\ }{\bf B42 (Proc.\ Suppl.)}\\(1995), 231;\\
		 L.\ Del Debbio, A.\  Di Giacomo and G.\ Paffuti and P.\
		 Pieri, {\em Phys.\ Lett.\ }{\bf B355}(1995), 255;\\
		 L.\ Del Debbio, A.\  Di Giacomo and G.\ Paffuti and P.\ Pieri,
		 {\em Nucl.\ Phys.\ }{\bf B42 (Proc. Suppl.)}(1995), 234;\\
		 T.\ Suzuki, S.\ Ilyar, Y.\ Matsubara, T.\ Okuda and K.\
		 Yotsuji,
		 {\em  Nucl.\ Phys.\ }{\bf B42 (Proc.\ Suppl.)}(1995), 481;\\
		 D.\ Stack, S.\ D.\ Neiman and  R.\ J.\ Wensley,
		 {\em Phys.\ Rev.\ }{\bf D50}(1994), 3399;\\
		 S.\ Ejiri, S.\ Kitahara, Y.\ Matsubara and T.\ Suzuki, {\em
		 Nucl.\ Phys.\ \ }{\bf B42 (Proc. Suppl.)} (1995), 481;\\
		 O.\ Miyamura and  S.\ Origuchi, QCD Monopoles and Chiral
		 Symmetry Breaking on SU(2) Lattices, preprint hep-lat/9508015.

\bibitem{MAP} G.\ 't Hooft, {\em Nucl.\ Phys.\ }{\bf B190}(1981), 455;\\
		 A.\ S.\ Kronfeld, G.\ Schierholz and U.\ Wiese,
		 {\em Nucl.\ Phys.\ }{\bf B 293}(1987), 461.

\bibitem{LNT} F.\ Lenz, H.\ W.\ L.\ Naus and M.\ Thies, {\em Ann.\ Phys.\
 		}(NY)\ {\bf 233}(1994), 317.

\bibitem{GJ} J.\ Goldstone and R.\ Jackiw, {\em Phys.\ Lett.\ }{\bf B74}(1978),
                 81;\\
		V.\ Baluni and B.\ Grossman, {\em Phys.\ Lett.}{\bf B78}(1978),
		 226;\\
		Y.\ A.\ Simonov, {\em Sov.\ J.\ Nucl.\ Phys.\ }{\bf 41}(1985),
                 835; 1014.

\bibitem{FNP} V.\ A.\ Franke, Y.\ V.\ Novozhilov and E.\ V.\ Prokhvatilov, {\em
                Lett.\ Math.\ Phys.\ }{\bf 5}(1981), 239.

\bibitem{Palumbo} F.\ Palumbo, {\em Phys.\ Lett.\ }{\bf B243}(1990), 109.

\bibitem{Yabuki} H.\ Yabuki, {\em Phys.\ Lett.\ }{\bf B231}(1989), 271.

\bibitem{Langetc} E.\ Langmann, M.\ Salmhover and A.\ Kovner, {\em
		 Mod.\ Phys.\ Lett.\ }{\bf A9}(1994), 2913.

\bibitem{BW} C.\ W.\ Bernard and E.\ J.\ Weinberg, {\em Phys.\ Rev.\ }{\bf
 		D15}(1977), 3656.

\bibitem{Singer} I.\ M.\ Singer, {\em Comm.\ Math.\ Phys.\ }{\bf 60}(1978), 7.

\bibitem{Gribov} V.\ N.\ Gribov, {\em  Nucl.\ Phys.\ }{\bf B139}(1978), 1.

\bibitem{Nakahara} M.\ Nakahara, Geometry, Topology and Physics, Adam Hilger
		 Graduate Student Series in Physics (1990), and references
		 therein.

\bibitem{JMR} R.\ Jackiw, I.\ Muzinich and C.\ Rebbi, {\em Phys.\ Rev.\ }{\bf
		 D17}(1978), 1576.

\bibitem{Cherno} M.\ N.\ Chernodub and F.\ V.\ Gubarev, Instantons and
                Monopoles in Maximal Abelian Projection of SU(2) Gluodynamics,
		 preprint ITEP-95-34, hep-th/9506026.

\bibitem{LNOT} F.\ Lenz, H.\ W.\ L.\ Naus, K.\ Ohta and M.\ Thies,
		 {\em Ann.\ Phys.\ }(NY)\ {\bf 233}(1994), 17.

\bibitem{tHo} G.\ 't Hooft, {\em Nucl.\ Phys.\ }{\bf B153}(1979), 141.

\bibitem{Lenzoffs} F.\ Lenz, E.\ J.\ Moniz and M.\ Thies, {\em Ann.\ Phys.\ }
                 (NY)\ {\bf 242}(1995), 429;\\
 		 F.\ Lenz, M.\ Shifman and M.\ Thies, {\em Phys.\ Rev.\ }{\bf
 		 D51}(1995), 7060;\\
		 D.\ Stoll and M.\ Thies, Higgs Mechanism and Symmetry Breaking
		 without Redundant Variables, preprint FAU-TP 3-95/3,
		 hep-ph/9504068.

\bibitem{hgrie} H.\ W.\ Grie\3hammer, PhD.\ Thesis, to be published.

\bibitem{GPY} D.\ J.\ Gross, R.\ D.\ Pisarski and L.\ G.\ Yaffe, {\em
		 Rev.\ Mod.\ Phys.\ }{\bf 53}(1981), 43.

\bibitem{JRCDG} R.\ Jackiw and C.\ Rebbi, {\em Phys.\ Rev.\ Lett.\
		}{\bf 37}(1976), 172; \\
		C.\ G.\ Callan, R.\ F.\ Dashen and D.\ J.\ Gross, {\em Phys.\
		 Lett.\ }{\bf 63B}(1976), 334.

\bibitem{Bala} A.\ P.\ Balachandran, Gauge Symmetries, Topology and
		  Quantization,
		  Lectures given at Miniworkshop on Methods of Electronic
                  Structure Calculations (part of the 1992 Research Workshop
                  in Condensed Matter, Atomic and Molecular Physics), Trieste,
                  Italy, preprint SU-4240-506, hep-th/9210111;\\
 		 A.\ P.\ Balachandran, G.\ Bimonte, K.\ S.\ Gupta and A.\
		 Stern, {\em Int.\ J.\ \-Mod.\-\ Phys.\ \-}{\bf A7}(1992),
		 4655.

\end{thebibliography}
\end{document}